\documentclass{article}
\usepackage{graphics}
\usepackage{dcolumn}
\usepackage{bm}

\begin{document}

\title{Quantum geons}
\author{Richard A. Sones\footnote{rasones@cs.com}}
\date{\today}
%******************************************************************
\maketitle
%******************************************************************
\begin{abstract}

We describe 
solutions of the Klein-Gordon equation which
are spherically symmetric
and localized, and may be regarded
as massive particles without charge or spin.
The proposed model, which is based on the action
for a complex scalar field minimally coupled to
the electromagnetic and gravitational fields, 
contains no adjustable parameters
and predicts five particle species with masses
of the order of the Planck mass.
These particles appear
to be candidates for dark matter.
\end{abstract}
%******************************************************************
\section{Introduction}

A number of models describing particles as solitons---stable
self-bound concentrations of field energy---have been proposed.
In an early paper Rosen~\cite{rosen39} obtained 
soliton solutions from the interaction of
a complex scalar field and the electromagnetic field.
Later Cooperstock and Rosen~\cite{cooperstock89} obtained
soliton solutions by coupling a complex scalar
field to both electromagnetism and gravity.
In the meantime Wheeler~\cite{wheeler55}, using an entirely
different theoretical framework, obtained soliton solutions
(which he dubbed geons)
from a purely classical model of electromagnetism coupled to
gravity. More recently Moroz, Penrose and Tod~\cite{moroz98}
obtained soliton solutions
of the Schr\"{o}dinger-Newton equations.
See \cite{poncedeleon04,bohun99,bodurov98,anderson97}
and references therein for a window into the literature.

The model presented here is similar to that of Cooperstock and
Rosen.\footnote{Our Eqs.\ (\ref{eq:1})--(\ref{eq:4})
are equivalent to Eqs.\ (5.3)--(5.6) of \cite{cooperstock89}.} 
What differs is the treatment of
boundary conditions and the introduction of certain constraints.
Cooperstock and
Rosen require all fields to be finite
and continuous at the center of the particle and they integrate the field equations
outward. Parameters are adjusted until
the asymptotic wave function vanishes. The resulting model contains several
adjustable parameters.

In contrast, we impose asymptotic boundary conditions 
and integrate
the field equations inward, toward the center of the particle.
And we require the asymptotically measured mass and
charge to equal the mass and charge parameters of the action.
The resulting model contains no adjustable parameters.

In Section \ref{sec:fieldequations} we write down the
action for a complex scalar field
minimally coupled to the electromagnetic and gravitational fields,
and use the principle of stationary action
to derive the corresponding field equations 
(the Klein-Gordon, Einstein and Maxwell equations).
Then we propose a stationary spherically symmetric trial solution 
for the metric tensor, wave function and vector potential and
substitute it into the field equations. We obtain
a system of four coupled nonlinear
differential equations, as well as two integral constraints
related to the conservation of charge and energy.
In Section \ref{sec:solution} we search for localized
particle-like solutions. After calculating
the asymptotic behavior of the equations and imposing
asymptotic boundary conditions, we are left with
a single adjustable parameter---the system electric charge.
By again appealing to the principle of stationary action
we find that the charge must vanish. Hence, we arrive at an eigensystem
with no adjustable parameters.
Using numerical methods we calculate the eigenmodes and
find five massive particle species without charge or spin.
We name these particles {\em quantum geons} and show that
the spacetime curvature diverges at the center of each geon.
In Section~\ref{sec:singularity} we explore the
implications of this singularity and argue that it is benign.
In Section~\ref{sec:discussion} we discuss our results.
%******************************************************************
\section{\label{sec:fieldequations}Field equations}

Consider the action
\begin{equation}
S = \frac{1}{16 \pi} \int _V \left[ {\sf R} - {\sf F}^{\mu \nu}{\sf F}_{\mu \nu}
    + {\textstyle \frac{1}{2}} \! \left( \, \overline{{\Psi}^{:\mu}} {\Psi}_{:\mu} 
    + {\Psi}^{:\mu} \overline{{\Psi}_{:\mu}} \, \right)
    - M^2 \overline{\Psi} \Psi
    \right] \surd \, d^{4} \! x
\label{eq:S}
\end{equation}
for a complex scalar field $\Psi$ within a four-dimensional volume
$V$ with spacetime curvature ${\sf R}$, where
a colon denotes the generalized covariant derivative \cite[p. 167]{fulling96}
\begin{equation}
{\Psi}_{:\mu} \equiv {\Psi}_{;\mu} + i Q {\sf A}_{\mu} \Psi \, ,
\end{equation}
a semicolon denotes the covariant derivative of general relativity,
an overline denotes complex conjugation,
$Q$ is electric charge,
${\sf A}_{\mu}$ is the electromagnetic vector potential,
\begin{equation}
{\sf F}_{\mu \nu} \equiv {\sf A}_{\mu ; \nu} - {\sf A}_{\nu ; \mu}
\end{equation}
is the electromagnetic field tensor
and $M$ is mass (assumed positive).
All quantities are expressed in natural units where
the speed of light, Planck's reduced constant ($\hbar$), Newton's
gravitational constant and Coulomb's electrostatic constant are unity.
Our notation and conventions are detailed in Appendix \ref{sec:Notation}.

Make small arbitrary variations $\delta S$ 
in the action by making
small arbitrary variations $\delta {\sf g}_{\alpha \beta}$,
$\delta {\sf A}_\mu$,
$\delta \Psi$ and $\delta \overline{\Psi}$ in the fields, 
and neglect boundary terms arising from integration by parts.
Then impose the condition $\delta S = 0$ for stationary action. 
The terms proportional to $\delta \Psi$ and $\delta \overline{\Psi}$
yield the Klein-Gordon equation
\begin{equation}
{{\Psi}^{:\mu}}_{:\mu} + M^2 \Psi = 0 
\label{eq:wave}
\end{equation}
and its complex conjugate.
The terms proportional to $\delta {\sf g}_{\alpha \beta}$
yield the Einstein equations
\begin{equation}
{\sf G}^{\alpha \beta} = -8 \pi {\sf T}^{\alpha \beta} \, ,
\label{eq:einstein}
\end{equation}
where
\begin{equation}
{\sf G}^{\alpha \beta} \equiv {\sf R}^{\alpha \beta} 
                              - {\textstyle \frac{1}{2}} {\sf g}^{\alpha \beta} {\sf R}
\end{equation}
is the Einstein tensor and
\begin{eqnarray}
{\sf T}^{\alpha \beta} &\equiv& \frac{1}{16 \pi}
                                \left( \overline{{\Psi}^{: \alpha}} {\Psi}^{:\beta} +
                                \Psi ^{: \alpha} \overline{{\Psi}^{:\beta}}
                                - {\sf g}^{\alpha \beta} \overline{{\Psi}^{:\mu}} \Psi _{: \mu}
                                + M^2 {\sf g}^{\alpha \beta} \overline{\Psi} \Psi \right) \nonumber \\
                            & & \mbox{} - \frac{1}{4 \pi} 
                                \left( {{\sf F}^{\alpha}}_{\mu} {\sf F}^{\beta \mu}
                                - {\textstyle \frac{1}{4}} \, {\sf g}^{\alpha \beta} 
                                {\sf F}^{\mu \nu} {\sf F}_{\mu \nu} \right)
\end{eqnarray}
is the energy-momentum tensor.
The terms proportional to $\delta {\sf A}_{\mu}$ yield the inhomogeneous Maxwell
equations
\begin{equation}
{{ \sf F}^{\mu \nu}}_{;\nu} = 4 \pi { \sf J}^{\mu} \, ,
\label{eq:inhomoMax}
\end{equation}
where
\begin{equation}
{ \sf J}^{\mu} \equiv \frac{iQ}{16 \pi}
                     \left( \overline{\Psi} {\Psi}^{:\mu}
                     - \overline{{\Psi}^{:\mu}} {\Psi} \right)
\end{equation}
is the electromagnetic current vector.
The homogeneous Maxwell equations
\begin{equation}
{ \sf F}_{\mu \nu ;\alpha} +
{ \sf F}_{\alpha \mu ;\nu} +
{\sf F}_{\nu \alpha ;\mu} = 0
\end{equation}
follow immediately from the definition of ${\sf F}_{\mu \nu}$.

Let the volume $V$ in Eq.\ (\ref{eq:S}) include
all of space over some arbitrary
time interval $T$.
We will later impose boundary conditions which require
${\sf R}$, ${\sf A}_{\mu}$ and $\Psi$ to vanish asymptotically.
Thus we are justified in neglecting boundary terms
when integrating by parts 
in the expression for $\delta S$.

When Eqs.\ (\ref{eq:wave}), (\ref{eq:einstein})
and (\ref{eq:inhomoMax})
are satisfied the action is stationary and
the scalar curvature is
\begin{equation}
{\sf R} = 8 \pi {\sf T}
        = 2 M^2 \overline{\Psi} \Psi
          - \overline{{\Psi}^{:\mu}} {\Psi}_{:\mu} \, ,
\end{equation}
where ${\sf T} \equiv {\sf T}_{\mu}{}^{\mu}$.
Substituting for ${\sf R}$ in Eq.\ (\ref{eq:S}) gives
\begin{equation}
S = \frac{1}{16 \pi} \int _V \left( M^2 \overline{\Psi} \Psi
    - {\sf F}^{\mu \nu} {\sf F}_{\mu \nu}
    \right) \surd \, d^{4} \! x
\label{eq:statS}
\end{equation}
as the stationary value of the action.
%******************************************************************
\subsection{Trial solution}
%See Geon166.nb.

Our goal is to solve Eqs.\ (\ref{eq:wave}), (\ref{eq:einstein})
and (\ref{eq:inhomoMax})
for the fields ${\sf g}_{\mu \nu}$, $\Psi$ and ${\sf A}_{\mu}$.
In order to make the calculations tractable
we consider a static spherically symmetric spacetime
with spherical polar coordinates $x^{0}=t$,
$x^{1}=r$, $x^{2}=\theta$ and $x^{3}=\phi$,
and metric tensor 
\begin{equation}
{\sf g}_{\mu \nu} = \left( 
\begin{array}{cccc}
u      & 0       & 0    & 0 \\
0      & -v      & 0    & 0 \\
0      & 0       & -r^2 & 0 \\
0      & 0       & 0    & -r^2 \sin ^2 \! \theta
\end{array}
\right) \, ,
\label{eq:g}
\end{equation}
where $u$ and $v$ are real functions of $r$.
The corresponding spherically symmetric wave function and
vector potential are assumed to have the forms
\begin{equation}
\Psi = R \, e^{-i \omega t}
\label{eq:Psi}
\end{equation}
and
\begin{equation}
Q {\sf A}_{\mu} = \left(
\begin{array}{c}
\Phi \\
0    \\
0    \\
0
\end{array} \right) \, ,
\label{eq:A}
\end{equation}
where $R$ and $\Phi$ are real functions of $r$ and
$\omega$ is a real constant.

Expressions (\ref{eq:g}), (\ref{eq:Psi}) and (\ref{eq:A})
for ${\sf g}_{\mu \nu}$, 
$\Psi$ and ${\sf A}_{\mu}$ comprise a trial solution
of field equations (\ref{eq:wave}), (\ref{eq:einstein})
and (\ref{eq:inhomoMax}) and
allow us to calculate all quantities of interest 
in terms of $R$, $u$, $v$ and $\Phi$.
We will show that the field equations
reduce to a set of four
coupled nonlinear differential equations in
$R$, $u$, $v$ and $\Phi$. Then we will solve
these equations subject to physically
plausible boundary conditions and constraints.
The fact that the field equations are, indeed, satisfied
by the assumed trial solution is its ultimate justification.
%******************************************************************
\subsection{Einstein, energy-momentum and electromagnetic tensors}

Given trial solution (\ref{eq:g}), (\ref{eq:Psi})
and (\ref{eq:A}) we can calculate the
quantities in field equations (\ref{eq:wave}), (\ref{eq:einstein})
and (\ref{eq:inhomoMax}).
The nonzero elements of the Einstein and energy-momentum tensors
are
\begin{eqnarray}
{{\sf G}_{0}}^{0} &=& -\frac{v'}{rv^2} - \frac{1}{r^2} + \frac{1}{r^{2}v} \nonumber \\
{{\sf G}_{1}}^{1} &=&  \frac{u'}{ruv}  - \frac{1}{r^2} + \frac{1}{r^{2}v} \nonumber \\
{{\sf G}_{2}}^{2} &=&  \frac{u''}{2uv} - 
\frac{\left( u' \right)^{2}}{4u^2v} - \frac{u' v'}{4uv^2}
+ \frac{u'}{2ruv} - \frac{v'}{2rv^2} \nonumber \\
{{\sf G}_{3}}^{3} &=& {{\sf G}_{2}}^{2}
\end{eqnarray}
and
\begin{eqnarray}
{{\sf T}_{0}}^{0} &=& \frac{M^2 R^2}{16 \pi}
                      + \frac{\left( \omega - \Phi \right)^2 R^2}{16 \pi u} 
                      + \frac{\left( R' \right) ^2}{16 \pi v}
                      + \frac{\left( \Phi ' \right) ^2}{8 \pi Q^2 uv} 
                      \nonumber \\
{{\sf T}_{1}}^{1} &=& \frac{M^2 R^2}{16 \pi}
                      - \frac{\left( \omega - \Phi \right)^2 R^2}{16 \pi u} 
                      - \frac{\left( R' \right) ^2}{16 \pi v}
                      + \frac{\left( \Phi ' \right) ^2}{8 \pi Q^2 uv} 
                      \nonumber \\
{{\sf T}_{2}}^{2} &=& \frac{M^2 R^2}{16 \pi}
                      - \frac{\left( \omega - \Phi \right)^2 R^2}{16 \pi u} 
                      + \frac{\left( R' \right) ^2}{16 \pi v}
                      - \frac{\left( \Phi ' \right) ^2}{8 \pi Q^2 uv} 
                      \nonumber \\
{{\sf T}_{3}}^{3} &=& {{\sf T}_{2}}^{2} \, ,
\label{eq:Txx}
\end{eqnarray}
where primes denote differentiation with respect to $r$.
The contraction of the energy-momentum tensor is
\begin{equation}
{\sf T}
= \frac{M^2 R^2}{4 \pi}
  - \frac{\left( \omega - \Phi \right)^2 R^2}{8 \pi u} 
  + \frac{\left( R' \right) ^2}{8 \pi v} \, .
\label{eq:T}
\end{equation}
The nonzero elements of the electromagnetic field tensor are
\begin{equation}
Q {\sf F}_{01} = -Q {\sf F}_{10} = \Phi '
\end{equation}
and the only nonzero element of the 
electromagnetic current vector is
\begin{equation}
{\sf J}^{0} = \frac{Q \left( \omega - \Phi \right) R^2}{8 \pi u} \, .
\label{eq:J0}
\end{equation}
We also note that
\begin{equation}
Q^2 {\sf F}^{\mu \nu} {\sf F}_{\mu \nu} = 
- \frac{2 \left( \Phi ' \right) ^2}{uv} \, .
\label{eq:FF}
\end{equation}
%******************************************************************
\subsection{Differential equations}

The Klein-Gordon equation (\ref{eq:wave}) becomes
\begin{equation}
R'' + \left( \frac{2}{r} + \frac{u'}{2u} - \frac{v'}{2v} \right) R'
+ \left[ \frac{\left( \omega - \Phi \right)^2}{u} - M^2 \right] vR = 0 \, .
\label{eq:1}
\end{equation}
There are four Einstein equations (\ref{eq:einstein}), one for
each nonzero component of ${{\sf G}_{\mu}}^{\nu}$.
However, the equations corresponding to ${{\sf G}_{2}}^{2}$ and
${{\sf G}_{3}}^{3}$ are identical. Furthermore, 
they can be derived by combining
the ${{\sf G}_{0}}^{0}$, 
${{\sf G}_{1}}^{1}$ and Klein-Gordon equations.
So we retain the ${{\sf G}_{0}}^{0}$, 
${{\sf G}_{1}}^{1}$ and Klein-Gordon equations, and discard the 
${{\sf G}_{2}}^{2}$ and ${{\sf G}_{3}}^{3}$ equations.
It will prove convenient to work with the equations obtained by
adding and subtracting the ${{\sf G}_{0}}^{0}$ 
and ${{\sf G}_{1}}^{1}$ equations:
\begin{equation}
\frac{u'}{2u} - \frac{v'}{2v} + \frac{1 - v}{r} =
- rv \left[ \frac{ M^2 R^2 }{2}
+ \frac{\left( \Phi ' \right) ^2}{Q^2 uv} \right]
\label{eq:2}
\end{equation}
and
\begin{equation}
\frac{u'}{2u} + \frac{v'}{2v} =
\frac{rv}{2} \left[ \frac{\left( \omega - \Phi \right)^2 R^2}{u} 
+ \frac{ \left( R' \right) ^2}{v} \right] \, .
\label{eq:3}
\end{equation}
The inhomogeneous Maxwell equations (\ref{eq:inhomoMax}) yield the Poisson equation
\begin{equation}
\Phi '' + \left( \frac{2}{r} - \frac{u'}{2u} - \frac{v'}{2v} \right) \Phi '
+ \frac{Q^2 R^2 v}{2} \, \left( \omega - \Phi \right) = 0 \, .
\label{eq:4}
\end{equation}

The four coupled nonlinear differential equations (\ref{eq:1})
through (\ref{eq:4}) are to be solved for the functions 
$R$, $u$, $v$ and $\Phi$ subject to physical constraints
and boundary conditions.
%******************************************************************
\subsection{\label{sec:conserved}Conserved quantities}

In this section we derive expressions for the
electric charge, energy, angular momentum 
and stationary action of our model.
%******************************************************************
\subsubsection{Charge}

The electromagnetic current vector satisfies the conservation law
\begin{equation}
{{\sf J}^{\mu}}_{;\mu} =
\left( {\sf J}^{\mu} \surd \right) _{,\mu} = 0 \, .
\end{equation}
The corresponding conserved quantity
\begin{equation}
Q = k_q \int_{\infty} {\sf J}^0 \surd \, dr \, d\theta \, d\phi
\label{eq:Q}
\end{equation}
is the electric charge,
where $k_q$ is a real constant (to be determined later).
Substitute ${\sf J}^0$ from Eq.\ (\ref{eq:J0}) and let
\begin{equation}
\surd = r^2 \sqrt{ \left| uv \right| } \, \sin \theta
\label{eq:surd}
\end{equation}
then integrate over $\theta$ and $\phi$ to get
\begin{equation}
1 = \frac{k_q}{2} \int_{0}^{\infty} \frac{r^2 \left( \omega - \Phi \right) R^2}{u} \sqrt{|uv|} \, dr \, .
\label{eq:Qnorm}
\end{equation}

Another expression for $Q$ can be obtained as follows.
Substituting from Eq.\ (\ref{eq:inhomoMax}) into 
Eq.\ (\ref{eq:Q}) and using the identity
\begin{equation}
{\sf F}^{\mu \nu}{}_{;\nu} \surd = 
\left( {\sf F}^{\mu \nu} \surd \right) _{,\nu}
\end{equation}
gives
\begin{equation}
Q = \frac{k_q}{4\pi} \int_{\infty} 
\left( {\sf F}^{0 \nu} \surd \right) _{, \nu} dr \, d\theta \, d\phi \, .
\end{equation}
For our static spherically symmetric metric the integrand is
\begin{equation}
\left( {\sf F}^{0 \nu} \surd \right) _{, \nu} =
\frac{d}{dr} \! \left( -\frac{\Phi '}{Quv} \surd \right) \, ,
\end{equation}
so
\begin{equation}
Q^2 = \frac{k_q}{4\pi} \int_{\infty} 
\frac{d}{dr} \! \left( -\frac{\Phi '}{uv} \surd \right) 
dr \, d\theta \, d\phi \, .
\end{equation}
Substituting expression (\ref{eq:surd}) for
$\surd$ then integrating over $\theta$ and $\phi$ gives
\begin{eqnarray}
Q^2 &=& {\int} _0 ^\infty \frac{d}{dr} \!
      \left( -\frac{k_q r^2 \Phi ' \sqrt{|uv|}}{uv} \right) \, dr \nonumber \\
  &=& \tilde{Q}^2(\infty) \, ,
\label{eq:Q3}
\end{eqnarray}
where
\begin{equation}
\tilde{Q}^2(r) \equiv -\frac{k_q r^2 \Phi ' \sqrt{|uv|}}{uv}
\label{eq:tildeQ}
\end{equation}
and we assume (to be justified later) $\tilde{Q}^2(0) = 0$.
The quantity $\tilde{Q}^2(r_b) - \tilde{Q}^2(r_a)$ is the 
square of the charge contained in the region between $r_a$ and $r_b$.
%******************************************************************
\subsubsection{\label{sec:energy}Energy}
%See p131 of ludvigsen99.

In our static spherically symmetric spacetime the vector field
\begin{equation}
{\sf \xi}^{\mu} = \left( \begin{array}{c}
                         1 \\
                         0 \\
                         0 \\
                         0                          \end{array}
                  \right)
\label{eq:xi}
\end{equation}
satisfies the Killing equation
\begin{equation}
{\sf \xi}^{\mu ; \nu} + {\sf \xi}^{\nu ; \mu} = 0 \, .
\end{equation}
If we let 
\begin{equation}
{\sf J}_{\xi}^{\mu} \equiv {{\sf \xi}^{\mu ; \nu}}_{; \nu}
\label{eq:Jxi}
\end{equation}
then the anti-symmetry of ${\sf \xi}^{\mu ; \nu}$ implies
\begin{equation}
{\sf J}_{\xi}^{\mu} \surd = \left( {\sf \xi}^{\mu ; \nu} \surd 
                            \right) _{, \nu}
\label{eq:gausslawxi}
\end{equation}
and
\begin{equation}
\left( {\sf J}_{\xi}^{\mu} \surd \right) _{, \mu} = 0 \, .
\end{equation}
Thus ${\sf J}_{\xi}^{\mu}$ is a conserved current associated with the time invariance
of the metric. The corresponding conserved quantity
\begin{equation}
M = k_{\xi} \int _{\infty} {\sf J}_{\xi}^{0} \surd \, dr \, d\theta \, d\phi
\label{eq:energy1}
\end{equation}
is the system mass (energy), 
where $k_{\xi}$ is a real constant (to be determined later).
Calculation of ${\sf J}_{\xi}^{\mu}$ from
Eqs.\ (\ref{eq:xi}) and (\ref{eq:Jxi}) for our
static spherically symmetric spacetime
gives
\begin{equation}
{\sf J}_{\xi}^{\mu} = \left( \begin{array}{c}
                             {{\sf R}_{0}}^{0} \\
                             0 \\
                             0 \\
                             0
                             \end{array}
                      \right) \, .
\end{equation}
But
\begin{equation}
{{\sf R}_{0}}^{0} = -8 \pi \left(
                     {{\sf T}_{0}}^{0} 
                     - {\textstyle \frac{1}{2}} {\sf T}
                     \right)
\end{equation}
so Eq.\ (\ref{eq:energy1}) becomes
\begin{equation}
M = -8 \pi k_{\xi} \int _{\infty} \left(
{{\sf T}_{0}}^{0} 
- {\textstyle \frac{1}{2}} {\sf T}
\right) 
\surd \, dr \, d\theta \, d\phi \, .
\label{eq:Komar}
\end{equation}
Substituting expressions (\ref{eq:Txx}), (\ref{eq:T})
and (\ref{eq:surd}) for ${{\sf T}_{0}}^{0}$, 
${\sf T}$ and $\surd$ into (\ref{eq:Komar}), 
then integrating over $\theta$ and $\phi$ 
gives
\begin{equation}
M = -4 \pi k_{\xi} \int_0^\infty r^2 \left[
    \frac{\left( \omega - \Phi \right)^2 R^2}{u}
    - \frac{M^2 R^2}{2} + \frac{\left( \Phi ' \right) ^2}{Q^2 uv} \right] \sqrt{|uv|} \, dr \, .
\label{eq:energy4}
\end{equation}

Another expression for $M$ can be obtained as follows.
Substituting from Eq.\ (\ref{eq:gausslawxi}) into Eq.\ (\ref{eq:energy1}) gives
\begin{equation}
M = k_{\xi} \int _{\infty} \left( \xi ^{0; \nu} \surd \right) _{,\nu}
\, dr \, d\theta \, d\phi \, .
\end{equation}
For our static spherically symmetric metric
the integrand is
\begin{equation} \left( \xi ^{0; \nu} \surd \right) _{,\nu} = \frac{d}{dr} \!
\left( -\frac{u'}{2uv} \surd \right) \, ,
\end{equation}
so
\begin{equation}
M = k_{\xi} \int _{\infty} 
\frac{d}{dr} \! \left( -\frac{u'}{2uv} \surd \right) 
\, dr \, d\theta \, d\phi \, .
\end{equation}
Substituting expression (\ref{eq:surd}) for
$\surd$ then integrating over $\theta$ and $\phi$ gives
\begin{eqnarray}
M &=& {\int} _0 ^\infty \frac{d}{dr} \!
      \left( -\frac{2 \pi k_{\xi} r^2 u'\sqrt{|uv|}}{uv} \right) \, dr \nonumber \\
  &=& \tilde{M}(\infty) \, ,
\label{eq:energy5}
\end{eqnarray}
where
\begin{equation}
\tilde{M}(r) \equiv -\frac{2 \pi k_{\xi} r^2 u' \sqrt{|uv|}}{uv}
\label{eq:tildeM}
\end{equation}
and we assume (to be justified later) $\tilde{M}(0) = 0$.
The quantity $\tilde{M}(r_b) - \tilde{M}(r_a)$ is the 
mass contained in the region between $r_a$ and $r_b$.
%******************************************************************
\subsubsection{Angular momentum}
%See p131 of ludvigsen99.

Based on the Killing field
\begin{equation}
{\sf \chi}^{\mu} = \left( \begin{array}{c}
                          0 \\
                          0 \\
                          0 \\
                          1
                          \end{array}
                   \right)
\end{equation}
and using an approach similar to that of Section \ref{sec:energy}
one can show that the system angular momentum is zero,
as expected for a spherically symmetric model.
%******************************************************************
\subsubsection{Action}

From Eqs.\ (\ref{eq:statS}), (\ref{eq:Psi}), (\ref{eq:FF})
and (\ref{eq:surd}) the stationary value of the action is
\begin{equation}
S = \frac{T}{2}\int_{0}^{\infty} r^2 \left[ \frac{ M^2 R^2}{2}
+ \frac{ \left( \Phi ' \right) ^2}{Q^2 uv} \right]
\sqrt{|uv|} \, dr \, .
\label{eq:S2}
\end{equation}
%******************************************************************
\section{\label{sec:solution}Solution of the field equations}

%******************************************************************
\subsection{\label{sec:infinity}Solution at infinity}

Our goal in this section is to determine
the asymptotic ($r \rightarrow \infty$) expressions for
the fields $R$, $u$, $v$ and $\Phi$.
We seek a particle-like solution 
with charge and energy distributions
localized near the origin and we impose the following
boundary conditions:
\begin{enumerate}
\item{The wave function $\Psi$
      and all components of the vector potential ${\sf A}_\mu$ 
      asymptotically vanish. \label{bc:vanish}}
\item{Spacetime is asymptotically flat and 
      local measurements by an asymptotic observer
      agree with the predictions of flat spacetime physics. 
      Specifically, the asymptotic behaviors
      of $u$ and $v$ reproduce Newtonian gravity,
      and the asymptotic expressions for $\Phi$ and $R$
      satisfy the flat spacetime Poisson and Klein-Gordon
      equations. \label{bc:flat}}
\end{enumerate}
These boundary conditions will guide us to candidate expressions
for the asymptotic fields. We will then verify by direct substitution
that these candidates asymptotically satisfy field equations (\ref{eq:1}) through (\ref{eq:4}).

The asymptotic expressions
\begin{eqnarray}
u &=& 1 - 2 M r^{-1} + Q^2 r^{-2} \nonumber \\
v &=& u^{-1}
\label{eq:asympuv}
\end{eqnarray}
correspond to the (Reissner-Nordstr\"{o}m) spacetime structure outside
a charged nonrotating black hole.
They yield flat spacetime
(Newtonian) gravity as $r \rightarrow \infty$, as required by
boundary condition \ref{bc:flat}.
They also imply
[see Eq.\ (\ref{eq:tildeM})]
\begin{equation}
\tilde{M}( \infty ) = -4 \pi k_{\xi} M \, .
\end{equation}
Comparing this with Eq.\ (\ref{eq:energy5}) gives
\begin{equation}
k_{\xi} = - \frac{1}{4 \pi} \, .
\end{equation}

The asymptotic expression
\begin{equation}
\Phi = Q^2 r^{-1}
\label{eq:asympPhi}
\end{equation}
follows from boundary conditions \ref{bc:vanish} and \ref{bc:flat} 
and the flat spacetime laws of electrostatics.
Expressions (\ref{eq:asympuv}) and (\ref{eq:asympPhi})
imply [see Eq.~(\ref{eq:tildeQ})]
\begin{equation}
\tilde{Q}^2( \infty ) = k_q Q^2 \, .
\end{equation}
Comparing this with Eq.\ (\ref{eq:Q3}) gives
\begin{equation}
k_q = 1 \, .
\end{equation}

The asymptotic form of $R$ follows from boundary condition \ref{bc:flat}
and the asymptotic flat spacetime Klein-Gordon equation
[obtained by letting
$u,v \rightarrow 1$ in Eq.\ (\ref{eq:1})]
\begin{equation}
R'' + \frac{2}{r} \, R'
+ \left[ \left( \omega - \Phi \right) ^2 - M^2 \right] R = 0 \, ,
\label{eq:1flat}
\end{equation}
where $\Phi$ is given by (\ref{eq:asympPhi}).
The solution which asymptotically vanishes (boundary condition \ref{bc:vanish})
is
%See Geon160.nb.
\begin{equation}
R = \frac{R_\infty \, e^{-kr}}{r^{1 + \sigma}} \, ,
\label{eq:asympR}
\end{equation}
where $R _\infty$ is a real constant, 
\begin{equation}
\sigma = \frac{Q^2 \omega}{k} \, ,
\label{eq:sigma}
\end{equation}
and
\begin{equation}
k = \sqrt{ M^2 - \omega ^2}
\label{eq:k}
\end{equation}
with $\omega ^2 < M^2$.

We now have asymptotic expressions for the fields which satisfy the
boundary conditions.
But we still need to verify that these expressions are consistent with
our field equations.
This is accomplished by
substituting the asymptotic expressions for $R$, $u$, $v$ and $\Phi$
into Eqs.\ (\ref{eq:1}) through (\ref{eq:4})
and verifying that the dominant terms satisfy the equations as $r \rightarrow \infty$.
This procedure reveals
that Eqs.\ (\ref{eq:2}) through (\ref{eq:4}) are, indeed,
asymptotically satisfied.
However, Eq.\ (\ref{eq:1}) is satisfied only if we let
(see Appendix \ref{sec:asympR})
\begin{equation}
\sigma = \frac{Q^2 \omega}{k} + \frac{M}{k} \left( M^2 - 2 \omega ^2 \right) \, ,
\label{eq:sigma2}
\end{equation}
which differs from (\ref{eq:sigma}).
In order to satisfy boundary condition \ref{bc:flat} we must
make (\ref{eq:sigma}) and (\ref{eq:sigma2}) consistent, which
requires
\begin{equation}
\omega = \pm M / \sqrt{2}
\label{eq:omega}
\end{equation}
and [from (\ref{eq:k})]
\begin{equation}
k = M / \sqrt{2} \, .
\label{eq:k2}
\end{equation}

In summary, based on the imposed boundary conditions
we expect the asymptotic solution to have the form
\begin{eqnarray}
R &=& R_{\infty} \, r^{-1 \mp Q^2} e^{- M r / \sqrt{2}} \nonumber \\
u &=& 1 - 2M r^{-1} + Q^2 r^{-2} \nonumber \\
v &=& u^{-1} \nonumber \\
\Phi &=& Q^2 r^{-1} \, .
\label{eq:asympRuvPhi}
\end{eqnarray}
%******************************************************************
\subsection{\label{sec:constraints}Physical constraints}

In addition to satisfying field equations (\ref{eq:1}) through (\ref{eq:4}) the solution must
also satisfy the charge and energy constraints of 
Section \ref{sec:conserved}.
Using results from Section \ref{sec:infinity}
in Eqs.\ (\ref{eq:Qnorm})
and (\ref{eq:energy4})
we obtain the constraints
\begin{equation}
1 = \frac{1}{2} \int_{0}^{\infty} r^2 \left( \pm \frac{M}{\sqrt{2}}
- \Phi \right) \frac{R^2}{u} \sqrt{|uv|} \, dr
\label{eq:constraint1}
\end{equation}
and
\begin{equation}
1 = \frac{1}{M} \int_0^\infty r^2 \left[
    {\left( \pm \frac{M}{\sqrt{2}} - \Phi \right) \!} ^2 \frac{R^2}{u}
    - \frac{M^2 R^2}{2}
    + \frac{\left( \Phi ' \right) ^2}{Q^2 uv} \right] \sqrt{|uv|} \, dr \, .
\label{eq:constraint2}
\end{equation}
%******************************************************************
\subsection{Solution strategy}

If we are given values of the three parameters
$Q$, $M$ and $R_\infty$
and a choice of sign in Eq.\ (\ref{eq:omega})
then the asymptotic fields
are fully defined by Eqs.\ (\ref{eq:asympRuvPhi}).
A complete solution can be obtained by
integrating field 
equations (\ref{eq:1}) through (\ref{eq:4}) from 
the asymptotic limit ($r \rightarrow \infty$)
down to the origin ($r \rightarrow 0$).
Since the two constraints of Section \ref{sec:constraints}
must be satisfied, only one of the three parameters is independent.
It will be convenient to take $Q$ as the independent
parameter and regard $M$ and $R_\infty$ as functions
of $Q$.

Once a solution is found for given $Q$ then the
action $S$ can be calculated from Eq.\ (\ref{eq:S2}).
We may regard the action as a function of $Q$.
The field equations, constraints and boundary
conditions actually depend on $Q^2$, not $Q$, so we may
write the action as $S(Q^2)$.
Solutions of physical interest have stationary action
so they must satisfy
\begin{equation}
\frac{\partial S}{\partial Q} = 
2 Q \, S' \! \! \left( Q^2 \right) = 0 \, .
\label{eq:dSdQ}
\end{equation}
%******************************************************************
\subsection{\label{sec:Q0}Solution for zero charge}

We will consider solutions with $Q=0$.
These solutions satisfy
Eq.\ (\ref{eq:dSdQ}) so they are of physical interest.

With $Q = 0$ the asymptotic fields (\ref{eq:asympRuvPhi})
become
\begin{eqnarray}
R &=& R_{\infty} \, r^{-1} e^{-Mr/\sqrt{2}} \nonumber \\
u &=& 1 - 2M r^{-1} \nonumber \\
v &=& u^{-1} \nonumber \\
\Phi &=& 0 \, .
\label{eq:asympRuvPhiQ0}
\end{eqnarray}
Field equation (\ref{eq:4}) becomes
\begin{equation}
\Phi '' + \left( \frac{2}{r} - \frac{u'}{2u} - \frac{v'}{2v} \right) \Phi ' = 0
\label{eq:4Q0}
\end{equation}
which, given the asymptotic expression for $\Phi$, yields
the trivial solution (valid everywhere)
\begin{equation}
\Phi = 0 \, .
\label{eq:PhiQ0}
\end{equation}
Then field equations (\ref{eq:1}) through (\ref{eq:3}) become
\begin{equation}
R'' + \left( \frac{2}{r} + \frac{u'}{2u} - \frac{v'}{2v} \right) R'
+ \left( \frac{1}{2u} - 1 \right) M^2 vR = 0 \, ,
\label{eq:1Q0}
\end{equation}
\begin{equation}
\frac{u'}{2u} - \frac{v'}{2v} + \frac{1 - v}{r} =
- \frac{M^2 rv R^2}{2}
\label{eq:2Q0}
\end{equation}
and
\begin{equation}
\frac{u'}{2u} + \frac{v'}{2v} =
\frac{rv}{2} \left[ \frac{M^2 R^2}{2u} 
+ \frac{\left( R' \right) ^2}{v} \right] \, ,
\label{eq:3Q0}
\end{equation}
and constraints (\ref{eq:constraint1}) and (\ref{eq:constraint2})
can be written
\begin{equation}
\Delta \equiv 2 \sqrt{2} \mp M \int_{0}^{\infty} \frac{r^2 R^2}{u} \sqrt{|uv|} \, dr = 0
\label{eq:constraint1Q0}
\end{equation}
and
\begin{equation}
\pm 2 \sqrt{2} - 2 - M \int_{0}^{\infty} 
r^2 R^2 \sqrt{|uv|} \, dr = 0 \, .
\label{eq:constraint2Q0}
\end{equation}
Since both $M$ and the integral in (\ref{eq:constraint2Q0}) are positive
this constraint can only be satisfied if we choose the upper sign.
So we need only consider the upper signs in 
Eqs.\ (\ref{eq:constraint1Q0}) and (\ref{eq:constraint2Q0}),
and [from Eq.\ (\ref{eq:omega})] $\omega = M/\sqrt{2}$.

Given values of $M$ and $R_{\infty}$, and starting 
with the asymptotic fields (\ref{eq:asympRuvPhiQ0}),
Eqs.\ (\ref{eq:1Q0}) through
(\ref{eq:3Q0}) can be numerically integrated all the way
to the origin. The values of $M$ and $R_{\infty}$ can be adjusted
until constraints (\ref{eq:constraint1Q0}) and (\ref{eq:constraint2Q0})
are satisfied.

\begin{figure}
\centering
\includegraphics{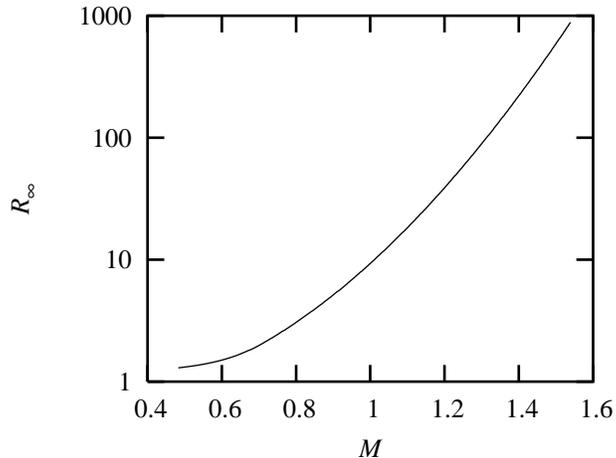}
\caption{\label{fig:Fig1}Locus in the $M$--$R_{\infty}$ plane which satisfies
constraint (\ref{eq:constraint2Q0}).}
\end{figure}

\begin{figure}
\centering
\includegraphics{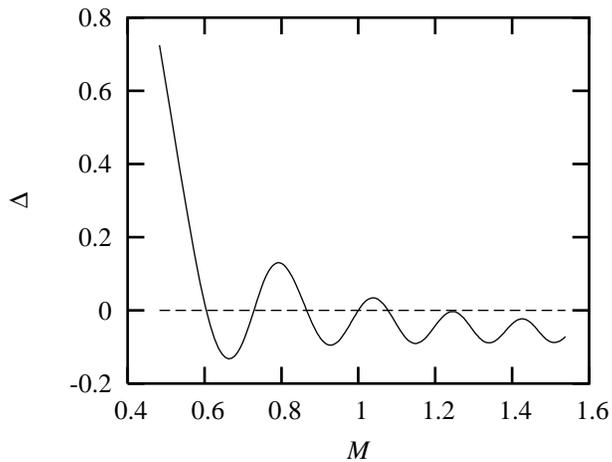}
\caption{\label{fig:Fig2}Left-hand side of constraint (\ref{eq:constraint1Q0})
versus mass
for the locus of points of Fig.\ \ref{fig:Fig1}. 
The five zero crossings
correspond to solutions of field equations (\ref{eq:1Q0})
through (\ref{eq:3Q0}) which satisfy boundary conditions (\ref{eq:asympRuvPhiQ0}) 
and constraints (\ref{eq:constraint1Q0}) and (\ref{eq:constraint2Q0}).
There are no zero crossings
in the neighborhood of $M = 1.25$.}
\end{figure}

Figure \ref{fig:Fig1} shows the locus in the $M$--$R_{\infty}$ plane
which satisfies constraint (\ref{eq:constraint2Q0}).
Figure \ref{fig:Fig2} plots $\Delta$ versus $M$ along that locus.
Since $\Delta = 0$ when constraint (\ref{eq:constraint1Q0}) is
satisfied the five zero crossings in Fig.~\ref{fig:Fig2} correspond
to the solutions we seek.
These solutions describe uncharged spinless
massive particles.
Because the masses appear spontaneously
as eigenvalues of a self-gravitating
quantum field (``mass without mass'') we
borrow terminology from Wheeler~\cite{wheeler55} and
dub these particles {\em quantum geons}.
We will index the five geons by the integer $n = 0,1,2,3,4$
in order of increasing mass.
%******************************************************************
%******************************************************************
%******************************************************************

\begin{figure}
\centering
\includegraphics{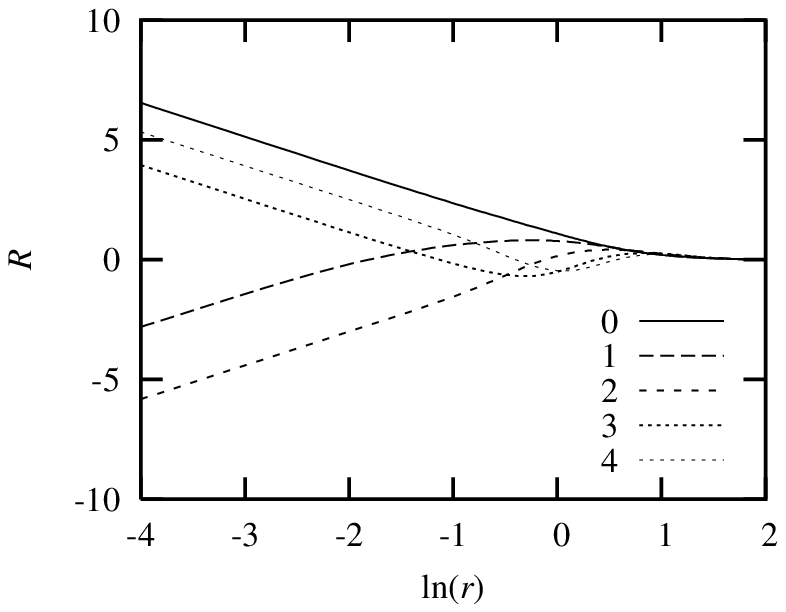}
\caption{\label{fig:Fig3}Radial wave function versus radius. 
Curves labeled by geon index $n$.}
\end{figure}

\begin{figure}
\centering
\includegraphics{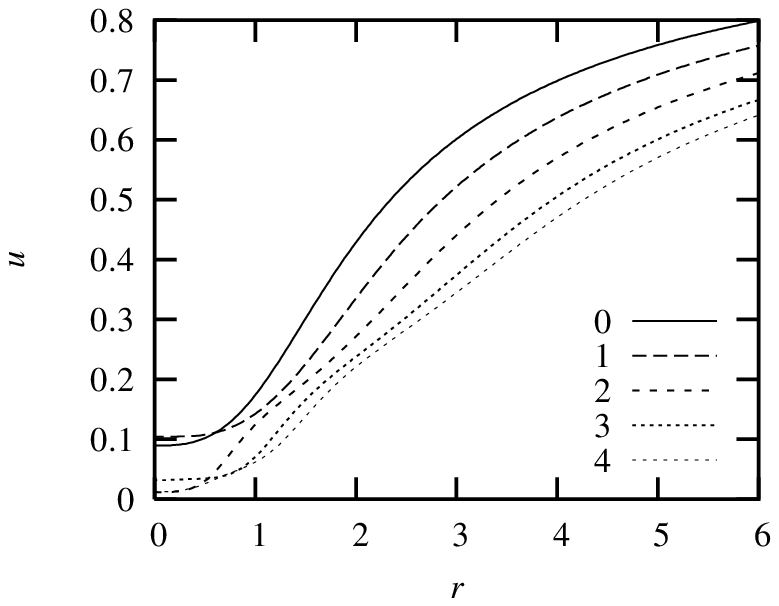}
\caption{\label{fig:Fig4}Metric function $u$ versus
radius. Curves labeled by geon index $n$.}
\end{figure}

\begin{figure}
\centering
\includegraphics{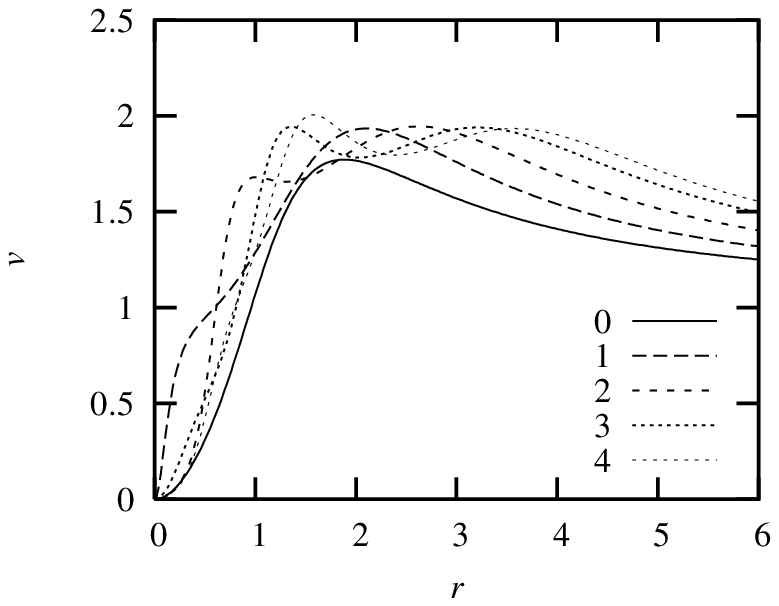}
\caption{\label{fig:Fig5}Metric function $v$ versus
radius. Curves labeled by geon index $n$.}
\end{figure}

\begin{figure}
\centering
\includegraphics{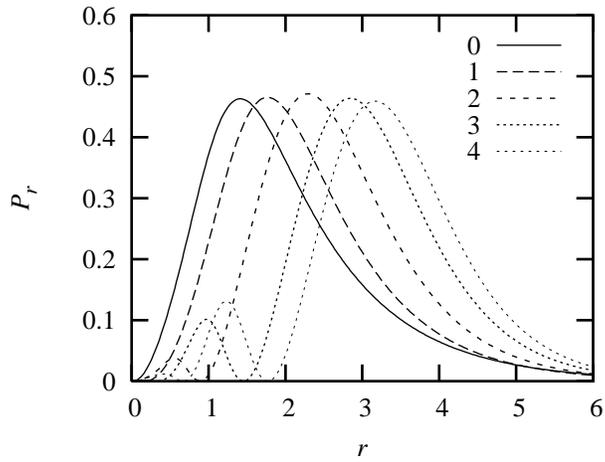}
\caption{\label{fig:Fig6}Radial probability density versus
radius. Curves labeled by geon index $n$.}
\end{figure}

Various features of the geons are plotted
in Figs.~\ref{fig:Fig3} through \ref{fig:Fig5}
and numerical characteristics are tabulated in
Table \ref{table:solutions}.
The probability density is
\begin{equation}
P \equiv \frac{\overline{\Psi} \Psi \surd}{\int _{\infty} \overline{\Psi} \Psi \surd \, dr \, d\theta \, d\phi} 
= N R^2 \surd \, ,
\end{equation}
where the normalization factor
\begin{equation}
N = \frac{M}{8 \pi \left( \sqrt{2} - 1 \right)}
\end{equation}
is determined from Eq.\ (\ref{eq:constraint2Q0}).
The radial probability density
\begin{equation}
P_r = 4 \pi N r^2 R^2 \sqrt{|uv|}
\end{equation}
is plotted in Fig.~\ref{fig:Fig6}.
The stationary value of the action calculated from
Eqs.~(\ref{eq:S2}) and (\ref{eq:constraint2Q0}) is
\begin{equation}
S = \frac{\left( \sqrt{2} - 1 \right) TM}{2}
\end{equation}
and the action per cycle ($T = 2\pi / \omega$)
is $(2 - \sqrt{2}) \pi$ for each of the five geons.
%******************************************************************
%******************************************************************
%******************************************************************
%******************************************************************
%See GeonMasses.mcd in Geon\Cpp\Geon004 folder.
\begin{table}
\centering
\caption{\label{table:solutions}Numerical characteristics of the geons.
Values of $M$ and $R_{\infty}$ are believed to be accurate to 
within $\pm 1$ of the least-significant
digit.}
\begin{tabular}{cclr}
\hline
\hline
$n$ & nodes in $R$ & \multicolumn{1}{c}{$\, \, M$} & \multicolumn{1}{l}{$\, \, R_{\infty}$} \\
\hline
$0$ & $0$ & $0.605$ & $ 1.51$     \\
$1$ & $1$ & $0.729$ & $ 2.22$     \\
$2$ & $1$ & $0.866$ & $ 4.26$     \\
$3$ & $2$ & $1.000$ & $ 9.3\;\;$     \\
$4$ & $2$ & $1.077$ & $15.6\;\;$     \\
\hline
\hline
\end{tabular}
\end{table}
%******************************************************************
%******************************************************************
\subsection{Solution at the origin}

The results of Section \ref{sec:Q0}
suggest that in the
neighborhood of the origin
\begin{eqnarray}
R &=& R_0 + R_{\ell} \ln r \nonumber \\
u &=& u_0 \nonumber \\
v &=& v_2 r^2 \, ,
\label{eq:Ruv_origin}
\end{eqnarray}
where $R_0$, $R_{\ell}$, $u_0$ and $v_2$ 
are real constants and $u_0,v_2 > 0$.
These expressions satisfy  
Eqs.\ (\ref{eq:1Q0}) through (\ref{eq:3Q0}) as $r \rightarrow 0$
provided
\begin{equation}
R_{\ell} = \pm \sqrt{2} \, .
\label{eq:Rell}
\end{equation}

Now that we know the behavior of the fields
as $r \rightarrow 0$ we are in a position to 
justify two earlier assumptions.
From expression~(\ref{eq:tildeQ}) for $\tilde{Q}^2(r)$
we obtain
\begin{equation}
\lim _{r \rightarrow 0} \tilde{Q}^2(r) = 
\lim _{r \rightarrow 0} \left[
- \frac{r \, \Phi' (r)}{\sqrt{u_0 v_2}} \right] = 0 \, ,
\label{eq:tildeQ2bc}
\end{equation}
justifying the assumption $\tilde{Q}^2(0) = 0$.
And from expression~(\ref{eq:tildeM}) for $\tilde{M}(r)$
we obtain
\begin{equation}
\lim _{r \rightarrow 0} \tilde{M}(r) =
\lim _{r \rightarrow 0} \left[
\frac{r \, u' (r)}{2\sqrt{u_0 v_2}} \right] = 0 \, ,
\label{eq:tildeMbc}
\end{equation}
justifying the assumption $\tilde{M}(0) = 0$.

The wave function $R$ diverges logarithmically at the origin.
This singularity is weak enough that
terms proportional to $R^2$ in the components of the energy-momentum 
tensor (\ref{eq:Txx}) are integrable.
But terms of the form $(R')^2/v$ diverge as $r^{-4}$
and are not integrable. These nonintegrable terms cancel out
of the expression $2{\sf T}_0{}^0 - {\sf T}$ so the
total energy is finite. 

Substituting expressions (\ref{eq:Ruv_origin})
into Eq.\ (\ref{eq:T}) for ${\sf T}$
and using ${\sf R} = 8 \pi {\sf T}$ gives
\begin{equation}
{\sf R} = \frac{2}{v_2 r^4}
\end{equation}
for the dominant behavior of the scalar curvature
as $r \rightarrow 0$.
Thus the scalar curvature diverges
at the center of each geon.
%******************************************************************
\section{\label{sec:singularity}Implications of infinite curvature}

Because the scalar curvature diverges at $r = 0$ it is not
possible to establish a locally flat coordinate system there.
So the locus $r = 0$ must be excluded from spacetime, leaving
a hole in the spacetime manifold.~\cite{hawking73}
The presence of this spacetime singularity 
raises the following questions:
\begin{enumerate}
\item{Do physical parameters (such as mass and charge) diverge?}
\item{Do arbitrary boundary conditions arise at the singularity?}
\item{Can particles and photons encounter the singularity and,
if so, what happens to them?} 
\end{enumerate}
Questions 1 and 2 have already been addressed:
the geon mass, charge, angular momentum and action
are all finite, and the wave function is normalizable;
and the conditions $\tilde{Q}^2(0)=0$ and $\tilde{M}(0)=0$ 
are not arbitrary boundary conditions at the 
singularity---they are consequences of the model [see Eqs.~(\ref{eq:tildeQ2bc})
and (\ref{eq:tildeMbc})].

The answer to question 3 is not so clear-cut.
In a classical (non-quantum) theory one considers
spacetime pathological if the world line of a 
freely falling test particle (a point particle of negligible mass)
does not exist after (or before) a finite interval of 
proper time. Such pathological spacetimes are
said to be timelike {\em geodesically incomplete}.~\cite{hawking73}
In a quantum theory, however, no real particle can
serve as a test particle in the
neighborhood of a singularity,
since such a particle would have to be so small
(short wavelength, large energy) that it would itself
dominate local spacetime structure.
Perhaps the concept of geodesic completeness can be extended
to the quantum realm~\cite{horowitz95},
but at the moment no consensus exists on how to do this.

So we will not be able to answer question 3 definitively.
But we will present a classical analysis which suggests the
singularity is benign.
In sections~\ref{sec:geodesics} and \ref{sec:geodesics2} we calculate 
timelike and null geodesics---the paths of freely falling particles and
photons---in the geon spacetime.
We will find that the geon core
is sufficiently repulsive to cloak the singularity and leave the
spacetime timelike and null geodesically complete.
%******************************************************************
\subsection{\label{sec:geodesics}Geodesic equations}

\begin{figure}
\centering
\includegraphics{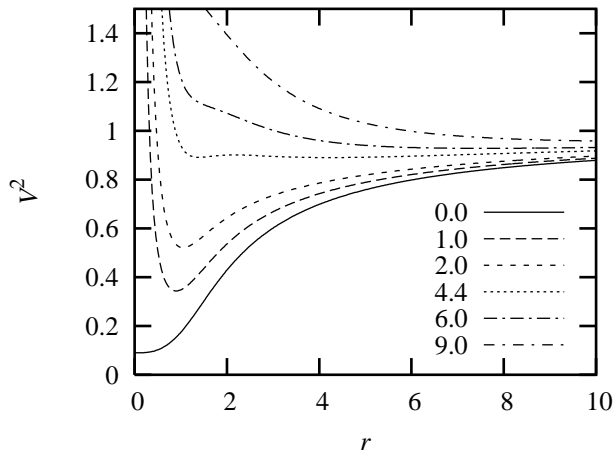}
\caption{\label{fig:timelike}Effective potential for radial motion along 
timelike geodesics in the spacetime
surrounding the $n = 0$ geon.
Curves labeled by $\ell ^2$. $V^2 \rightarrow 1$ as $r \rightarrow \infty$.}
\end{figure}

\begin{figure}
\centering
\includegraphics{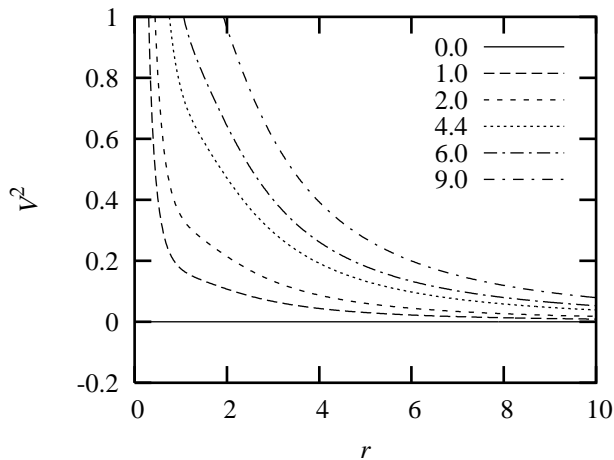}
\caption{\label{fig:null}Effective potential for radial motion along 
null geodesics in the spacetime
surrounding the $n = 0$ geon.
Curves labeled by $\ell ^2$. $V^2 \rightarrow 0$ as $r \rightarrow \infty$.}
\end{figure}

Consider moving along a geodesic
with velocity ${\sf v}^{\mu}$.
The trajectory obeys
\begin{equation}
\frac{d {\sf v}_{\alpha}}{ds} = 
{\textstyle \frac{1}{2}} {\sf g}_{\mu \nu , \alpha}
{\sf v}^{\mu} {\sf v}^{\nu}
\label{eq:geodesic}
\end{equation}
subject to the constraint
\begin{equation}
{\sf v}^{\mu} {\sf v}_{\mu} = \zeta \, ,
\label{eq:vnorm}
\end{equation}
where $\zeta = 0,1$ for null and timelike geodesics, and 
$s$ (proper time for a timelike geodesic) parameterizes the trajectory.
${\sf g}_{\mu \nu}$ is independent of $t$ and $\phi$ so
equation (\ref{eq:geodesic}) implies ${\sf v}_t$ and
${\sf v}_\phi$ are constant along the trajectory.
Given the spherical symmetry we can
(without loss of generality) consider trajectories confined to the
equatorial plane ($\theta = \pi / 2$, 
${\sf v}_\theta = {\sf v}^\theta = 0$).
Thus
\begin{equation}
{\sf v}_{\mu} = \left(
\begin{array}{c}
\varepsilon \\
-v \dot{r} \\
0 \\
- \ell \\
\end{array} \right)
\mbox{\hspace{0.15in} and \hspace{0.15in}}
{\sf v}^{\mu} = \left(
\begin{array}{c}
\varepsilon /u \\
\dot{r} \\
0 \\
\ell /r^{2} \\
\end{array} \right) \, ,
\label{eq:vdandvu}
\end{equation}
where $\dot{r} \equiv dr/ds$, and $\varepsilon$ and $\ell$ are real constants.
Then equation (\ref{eq:vnorm}) gives
\begin{equation}
uv \dot{r} ^2 =
\varepsilon ^2 - V^2 \, ,
\label{eq:geodesic2}
\end{equation}
where
\begin{equation}
V^2 \equiv u \left( \zeta + \frac{\ell ^2}{r^2} \right) \, .
\end{equation}
The evolution of $\phi$ is determined by
\begin{equation}
\dot{\phi} \equiv d\phi / ds = v^{\phi} = \ell / r^2 \, ,
\label{eq:phidot}
\end{equation}
and this equation and (\ref{eq:geodesic2}) completely determine the
trajectory.
Since $u,v \ge 0$ equation (\ref{eq:geodesic2})
can be satisfied only when $\varepsilon ^2 \ge V^2$, and
the condition $\varepsilon ^2 = V^2$
corresponds to a turning point of the radial motion.
We will refer to $V^2$ as the {\em effective potential} (for radial motion).
The effective potentials for timelike and null geodesics in the spacetime
surrounding the $n = 0$ geon
are plotted in Figs.~\ref{fig:timelike} and \ref{fig:null}.
%******************************************************************
%******************************************************************
%******************************************************************
\subsection{\label{sec:geodesics2}Geodesics near the singularity}

As $r \rightarrow 0$ the metric approaches
\begin{equation}
ds^2 = u_0 \, dt^2 - v_2 r^2 \, dr^2 - r^2 \, d\theta^2 - r^2 \, \sin^2 \! \theta \, d\phi^2 \, .
\label{eq:metricr0}
\end{equation} 
Using the results of section \ref{sec:geodesics}
we will explore the equatorial geodesics of this metric.

\begin{figure}
\centering
\includegraphics{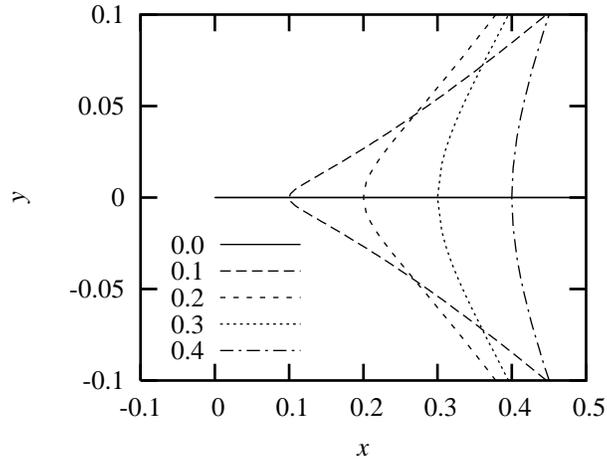}
\caption{\label{fig:singularity}Geodesics associated with
the metric (\ref{eq:metricr0}) near the center of a
geon. See Eqs.~(\ref{eq:xy}). Curves labeled by $r_0^*$.}
\end{figure}

Let $r_0$ represent the minimum value of $r$ along a geodesic.
By setting $\varepsilon ^2 = V^2$ we find
\begin{equation}
r_0 = \sqrt{\frac{u_0 \ell ^2}{\varepsilon ^2 - u_0 \zeta}} \, .
\end{equation}
The geodesic trajectory is determined by
\begin{equation}
\frac{dr}{d\phi} = \frac{\dot{r}}{\dot{\phi}} = \pm \frac{1}{\sqrt{v_2}}
\, \sqrt{\left( \frac{r}{r_0} \right) ^2 - 1}
\end{equation}
which can be integrated to give
\begin{equation}
\phi = \pm \, r_0^* \, \ln \! \! \left[ \frac{r^*}{r_0^*} + 
\sqrt{\left( \frac{r^*}{r_0^*} \right) ^2 - 1} \, \right] \, ,
\label{eq:phi}
\end{equation}
where we have introduced the scaled radial coordinate
$r^* = \sqrt{v_2} \, r$ and
chosen the constant of integration so
$\phi = 0$ when $r^* = r_0^*$.
Geodesics corresponding to (\ref{eq:phi}) are plotted in terms
of the Cartesian coordinates
\begin{eqnarray}
x &\equiv& r^* \cos \phi \nonumber \\
y &\equiv& r^* \sin \phi
\label{eq:xy}
\end{eqnarray}
in Fig.~\ref{fig:singularity}.
Test particles are attracted when $r^* \gg 1$ 
and repelled when $r^* \ll 1$.
In the limit of zero angular momentum ($\ell = r_0^* = 0$) the geodesic
is coincident with the positive $x$ axis, corresponding to
a test particle rebounding directly backwards.
%******************************************************************
%******************************************************************

\begin{figure}
\centering
\includegraphics{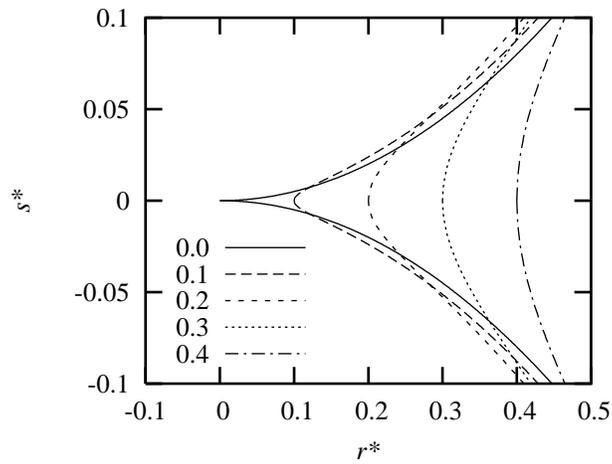}
\caption{\label{fig:singularity2}Parameter $s^*$
versus radius $r^*$ for geodesics associated with
the metric (\ref{eq:metricr0}) near the center of a
geon. Curves labeled by $r_0^*$.}
\end{figure}

Now let us calculate the behavior of
$s$ as a function of $r$ for particles and photons 
following the geodesics of Fig.~\ref{fig:singularity}.
From Eq.~(\ref{eq:geodesic2}) we obtain
\begin{equation}
\left( r \dot{r} \right) ^2 = \frac{\varepsilon ^2 - V^2}{u_0 v_2}
= \kappa ^2 \left[ 1 - \left( \frac{r_0}{r} \right) ^2 \right] \, ,
\label{eq:dr2ds}
\end{equation}
where
\begin{equation}
\kappa \equiv \sqrt{ \frac{\varepsilon ^2 - u_0 \zeta}{u_0 v_2} } \, .
\end{equation}
Integrating (\ref{eq:dr2ds}) we obtain
\begin{equation}
s^* = \pm \frac{(r_0^*)^2}{2} \left\{ \frac{r^*}{r_0^*}
\sqrt{ \left( \frac{r^*}{r_0^*} \right) ^2 - 1} +
\ln \! \! \left[ \frac{r^*}{r_0^*} + \sqrt{ \left( \frac{r^*}{r_0^*} \right) ^2 - 1} \, \right] \right\} \, ,
\label{eq:s}
\end{equation}
where we have introduced the scaled parameter $s^* = v_2 \kappa s$ and
chosen the constant of integration so
$s^* = 0$ when $r^* = r_0^*$.
Equation~(\ref{eq:s}) is plotted in Fig.~\ref{fig:singularity2}
for various values of $r_0^*$.
%******************************************************************
%******************************************************************

Figures~\ref{fig:singularity} and \ref{fig:singularity2}
suggest that the core of the geon behaves 
like a repulsive
potential barrier, effectively cloaking the singularity. 
Freely falling test particles with zero angular momentum
($\ell = r_0^* = 0$) do encounter the singularity,
but they rebound---their world lines do not vanish.
The spacetime is timelike and null geodesically complete.

One may object that any test particle encountering
the singularity will be destroyed by
infinite tidal forces. But keep in mind that
no real particle can serve as a test particle
at the Planck scale, so infinite tidal forces at the
singularity do not necessarily pose a physical conundrum.
%******************************************************************
\section{\label{sec:discussion}Discussion}

The use of boundary condition \ref{bc:flat} to force 
correspondence between
Eqs.\ (\ref{eq:sigma}) and (\ref{eq:sigma2}) can be 
motivated in the following way.
Consider two different localized physical systems A and B
with zero angular momentum and identical charges and masses.
Suppose A
is accurately described by the flat 
spacetime Klein-Gordon equation (\ref{eq:1flat})
while B is described by the curved spacetime equation (\ref{eq:1}).
Viewed from a sufficiently large distance each system appears as
a point particle with a wave function given by (\ref{eq:asympR}).
Using local measurements of the probability density $\overline{\Psi} \Psi$ an
asymptotic observer could determine $\sigma$ and thereby
discriminate between the particles, despite the fact that both
have identical charge, angular momentum and mass.
It is physically appealing to expect the asymptotic 
behavior of $\overline{\Psi} \Psi$
to be uniquely determined by these three parameters.
Indeed, this is what we typically mean by the word
``particle''.
By selecting $\omega$ to force $\sigma$ to be identical
for particles A and B we guarantee that the asymptotic
behavior of $\overline{\Psi} \Psi$ depends only on charge,
angular momentum and mass, and is independent of internal details.

At the Planck scale many physicists
expect classical notions of spacetime to fail---a concept
conveyed by the phrase (also coined by Wheeler) ``spacetime foam''.
A successful quantum theory of gravity would, it is thought, 
flesh out the details of spacetime foam and
erase the singularities associated with point particles.

The geon model developed here suggests a different perspective.
Classical spacetime is assumed valid at the Planck scale and
point particles are replaced by eignemodes of a quantum field. 
Singularities reminiscent of
point particles remain, but they do not disturb the
geodesic completeness of spacetime.
Multi-particle systems would, presumably, correspond to
multiple excitations of the geon
field \cite{rosen52,bodurov98} and all particle interactions
(including those involving wave-packet
reduction) would ultimately derive
from the action (\ref{eq:S}).
In short, the geon perspective replaces the
search for a ``quantum theory of gravity'' with the search for
a ``gravitational theory of quanta''.~\cite{bodurov98}

Some questions for future investigation come to mind:
\begin{enumerate}

\item{Do charged solutions of Eq.\ (\ref{eq:dSdQ}) exist?}

\item{Do solutions with nonzero angular momentum exist?
The spherically symmetric trial solution 
[Eqs.~(\ref{eq:g}), (\ref{eq:Psi})
and (\ref{eq:A})] could be
replaced by one with axial symmetry.
However, the trial
solution would now involve $\theta$ and $\phi$ as well as $r$ and $t$, 
and the metric tensor and vector potential would have additional
nonzero components, so the solution would be much more challenging.
Furthermore, it is not obvious what functional form 
should replace metric tensor (\ref{eq:g}).}

\item{What is the physical significance of the angular frequency 
$\omega = M / \sqrt{2}$?}

\item{What is the physical significance of the stationary value of the
action and the action per cycle?}

\item{The action (\ref{eq:S}) is, arguably, the simplest which
includes gravitation, electromagnetism and quantum mechanics.
But there is no compelling reason to regard it
as correct.~\cite[p. 144]{feynman95} It would be interesting to see how
other terms in the action (such as a cosmological constant or 
conformal coupling~\cite[p. 116]{fulling96}) affect geon solutions.}

\end{enumerate}

The geons described here
are far too massive to correspond to any known particle.
They would interact gravitationally with ordinary matter
so they appear to be candidates for dark matter.
A number of workers have considered Planck-mass particles
as dark matter (see \cite{chen04} and references therein)
but it is not clear whether such models can be successfully incorporated
into standard cosmology.
The density of dark matter within a galactic halo
is thought to be about $0.3 \, \mbox{GeV} \, \mbox{cm}^{-3}$ \cite{akimov01}
so, if all dark matter is composed of Plank-mass 
($1.2 \times 10^{19} \, \mbox{GeV}$) geons,
the local geon number density
is about $0.3 \times 10^{-19} \, \mbox{cm}^{-3}$.
This tiny density and weak coupling to ordinary matter
would make the detection of such 
particles difficult.
%******************************************************************
\appendix
%******************************************************************
\section{\label{sec:Notation}Notation and conventions}

Our notation follows Dirac \cite{dirac75}.
Greek indices take on the values $0, 1, 2, 3$ and repeated
indices are summed.
The spacetime coordinates are ${\sf x}^{\mu}$ with ${\sf x}^{0}=t$ and
the metric signature is $+$$-$$-$$-$.

The curvature tensor is
\begin{equation}
{{\sf R}^{\alpha}}_{\mu \nu \beta} \equiv
\mbox{} - {{\sf \Gamma} ^{\alpha}}_{\mu \nu , \beta}
+ {{\sf \Gamma} ^{\alpha}}_{\mu \beta , \nu}
\mbox{} - {{\sf \Gamma} ^{\sigma}}_{\mu \nu} {{\sf \Gamma} ^{\alpha}}_{\sigma \beta}
+ {{\sf \Gamma} ^{\sigma}}_{\mu \beta} {{\sf \Gamma} ^{\alpha}}_{\sigma \nu} \, ,
\end{equation}
the Ricci tensor is ${\sf R}_{\mu \nu} \equiv {{\sf R}^{\alpha}}_{\mu \nu \alpha}$
and the scalar curvature is ${\sf R} \equiv {{\sf R}_{\mu}}^{\mu}$,
where
\begin{equation}
{\sf \Gamma} _{\alpha \mu \nu} \equiv {\textstyle \frac{1}{2}} \left( {\sf g}_{\alpha \mu , \nu}
+ {\sf g}_{\alpha \nu , \mu} - {\sf g}_{\mu \nu , \alpha} \right)
\end{equation}
is the Christoffel symbol and a comma preceding 
some lower index $\mu$ denotes partial differentiation with respect to ${\sf x}^{\mu}$.
The metric tensor ${\sf g}_{\mu \nu}$ is symmetric, its contraction is
\begin{equation}
{\sf g}_{\mu \nu} {\sf g}^{\mu \nu} = {{\sf g}_{\mu}}^{\mu} = 4 \, ,
\end{equation}
and, for notational brevity,
\begin{equation}
\surd \equiv \sqrt{|\det \! \left( {\sf g}_{\mu \nu} \right)|} \, .
\end{equation}

A semicolon or colon preceding some lower index $\mu$ denotes
a covariant derivative with respect to ${\sf x}^{\mu}$.
A semicolon denotes the covariant derivative of general relativity.
For a scalar, covariant vector and contravariant vector
\begin{eqnarray}
\Psi _{; \mu} &\equiv& \Psi _{, \mu} \nonumber \\
{\sf V}_{\mu ; \nu} &\equiv& {\sf V}_{\mu , \nu}
- {{\sf \Gamma} ^{\alpha}}_{\mu \nu} {\sf V}_{\alpha} \nonumber \\
{{\sf V}^{\mu}}_{; \nu} &\equiv& {{\sf V}^{\mu}}_{, \nu} 
+ {{\sf \Gamma} ^{\mu}}_{\alpha \nu} {\sf V}^{\alpha}
\end{eqnarray}
and the procedure
generalizes in the usual way to tensors of any rank and 
mixture of covariant and contravariant indices.
A colon denotes the generalized covariant derivative
\begin{equation}
{\sf U}_{:\mu} \equiv 
{\sf U}_{;\mu} + 
iQ {\sf A}_{\mu} {\sf U}
\end{equation}
for an arbitrary tensor ${\sf U}$.
%******************************************************************
\section{\label{sec:asympR}Asymptotic wave function}

In the asymptotic regime [see Eqs.\ (\ref{eq:asympuv})
and (\ref{eq:asympPhi})]
\begin{equation}
\Phi = Q^2r^{-1} + O(r^{-2})
\end{equation}
and
\begin{equation}
u = 1 - 2Mr^{-1} + Q^2r^{-2} + O(r^{-3}) \, .
\end{equation}
By expanding $u^{-1}$ in terms of $r^{-1}$ as $r \rightarrow \infty$
we obtain
%See Geon159.nb.
\begin{equation}
v = 1 + 2Mr^{-1} + \left( 4M^2 -Q^2 \right) r^{-2} + O(r^{-3}) \, .
\end{equation}
Consider a trial solution for $R$ of the form
\begin{equation}
R = R_{\infty} \left[ 1 +
ar^{-1} + O(r^{-2}) \right] r^{-1-\sigma} e^{-kr} \, ,
\end{equation}
where $a$ is a real constant.
Substituting the above expressions into Eq.\ (\ref{eq:1})
and expanding in terms of $r^{-1}$ as $r \rightarrow \infty$
gives
%See Geon159.nb.
\begin{equation}
0 = \kappa +
\left[ \kappa a -
2M \left( M^2 - 2 \omega ^2 \right) - 2 Q^2 \omega + 2k \sigma \right] r^{-1}
+ O(r^{-2}) \, ,
\label{eq:1asymp}
\end{equation}
where
\begin{equation}
\kappa \equiv k^2 + \omega ^2 - M^2 \, .
\end{equation}
Equation (\ref{eq:1asymp}) must be satisfied term-by-term so
$\kappa = 0$ and we obtain Eqs.\ (\ref{eq:k}) and (\ref{eq:sigma2}).
%******************************************************************
%\bibliography{../geon.bib}

\begin{thebibliography}{10}

\bibitem{akimov01}
D.~Yu. Akimov.
\newblock Experimental methods for particle dark matter detection (review).
\newblock {\em Instrum. Exp. Tech. (Russia)}, 44(5):575--617, 2001.

\bibitem{anderson97}
Paul~R. Anderson and Dieter~R. Brill.
\newblock Gravitational geons revisited.
\newblock {\em Phys. Rev. D}, 56(8):4824--4833, 1997.

\bibitem{bodurov98}
Theodore Bodurov.
\newblock Complex {H}amiltonian evolution equations and field theory.
\newblock {\em J. Math. Phys.}, 39(11):5700--5715, 1998.

\bibitem{bohun99}
C.~Sean Bohun and F.~I. Cooperstock.
\newblock {D}irac-{M}axwell solitons.
\newblock {\em Phys. Rev. A}, 60(6):4291--4300, 1999.

\bibitem{chen04}
Pisen Chen.
\newblock {P}lanck-size black hole remnants as dark matter.
\newblock {\em Mod. Phys. Lett. A}, 19(13-16):1047--1054, 2004.

\bibitem{cooperstock89}
F.~I. Cooperstock and N.~Rosen.
\newblock A nonlinear gauge-invariant field theory of leptons.
\newblock {\em Int. J. Theor. Phys.}, 28(4):423--440, 1989.

\bibitem{dirac75}
P.~A.~M. Dirac.
\newblock {\em General Theory of Relativity}.
\newblock John Wiley \& Sons, 1975.

\bibitem{feynman95}
Richard~P. Feynman, Fernando~B. Morinigo, and William~G. Wagner.
\newblock {\em Feynman Lectures on Gravitation}.
\newblock Perseus Books, 1995.

\bibitem{fulling96}
Stephen~A. Fulling.
\newblock {\em Aspects of Quantum Field Theory in Curved Space-Time}.
\newblock Cambridge University Press, 1996.

\bibitem{hawking73}
S.~W. Hawking and G.~F.~R. Ellis.
\newblock {\em The large scale structure of space-time}.
\newblock Cambridge University Press, 1973.

\bibitem{horowitz95}
Gary~T. Horowitz and Donald Marolf.
\newblock Quantum probes of spacetime singularities.
\newblock {\em arXiv.org}, arXiv:gr-qc/9504028v3, 1995.

\bibitem{moroz98}
Irene~M. Moroz, Roger Penrose, and Paul Tod.
\newblock Spherically-symmetric solutions of the {S}chr{\"{o}}dinger-{N}ewton
  equations.
\newblock {\em Class. Quantum Grav.}, 15:2733--2742, 1998.

\bibitem{poncedeleon04}
J.~{Ponce de Leon}.
\newblock Electromagnetic mass-models in general relativity reexamined.
\newblock {\em Gen. Relativ. Gravit.}, 36(6):1453--1461, 2004.

\bibitem{rosen39}
N.~Rosen.
\newblock A field theory of elementary particles.
\newblock {\em Phys. Rev.}, 55:94--101, 1939.

\bibitem{rosen52}
Nathan Rosen and Herbert~B. Rosenstock.
\newblock The force between particles in a nonlinear field theory.
\newblock {\em Phys. Rev.}, 85(2):257--259, 1952.

\bibitem{wheeler55}
J.~A. Wheeler.
\newblock Geons.
\newblock {\em Phys. Rev.}, 97:511--536, 1955.

\end{thebibliography}

%******************************************************************
\end{document}